\documentclass[12pt]{iopart} \usepackage{graphicx,amssymb}
\expandafter\let\csname equation*\endcsname\relax
\expandafter\let\csname endequation*\endcsname\relax
\usepackage{amsmath}

\usepackage{color}
\usepackage{graphicx}
\usepackage{float}
\usepackage{subfigure}
\usepackage{hyperref}
\hypersetup{
    colorlinks=true,
    linkcolor=blue,
    filecolor=magenta,      
    urlcolor=cyan,
}
 
\usepackage{amsmath}
\usepackage{acronym}
\usepackage{multirow}
\usepackage{tabu}
\usepackage{tikz}
\usetikzlibrary{arrows,shapes,trees,decorations.pathreplacing}
\usepackage{import}
\usepackage{svn-multi}
\usepackage{lineno}

\begin{document}

\title[]{Distinguishing short
duration noise transients in LIGO data to improve
the PyCBC search for gravitational waves from
high mass binary black hole mergers.}
\author{Alexander H. Nitz$^1$, 
        }
        
\address{$^1$ Max-Planck-Institut f{\"u}r Gravitationsphysik,
         Albert-Einstein-Institut, D-30167 Hannover, Germany}
\date{\today}

\begin{abstract}
"Blip glitches" are a type of short duration transient noise in LIGO data. The cause for the majority of these is currently unknown. Short duration transient noise creates challenges for searches of the highest mass binary black hole systems, as standard methods of applying signal consistency, which look for consistency in the accumulated signal-to-noise of the candidate event, are unable to distinguish many blip glitches from
short duration gravitational-wave signals due to similarities in their time and frequency evolution. We demonstrate a straightforward method,
employed during Advanced LIGO's second observing run, including the period of joint observation with
the Virgo observatory, to separate the majority of this transient noise from potential gravitational-wave sources. This yields a $\sim 20\%$ improvement in the detection rate of high mass binary black hole mergers ($> 60 M_{\odot}$) for the PyCBC analysis.
\end{abstract}

\maketitle

\section{Introduction}
\label{s:intro}

Advanced LIGO has only recently completed its second observing run (O2). Including the first observing run, multiple binary black hole mergers have been reported~\cite{Abbott:2016blz,Abbott:2016nmj,Abbott:2017vtc,Abbott:2017oio}. Continued observation provides
the opportunity for additional detections, and to improve our understanding of the population
of binary black holes~\cite{TheLIGOScientific:2016wfe,TheLIGOScientific:2016pea}. Searches for gravitational waves from binary black hole mergers, and other compact object mergers, make use of 
matched filtering, which uses a waveform model~\cite{Nitz:2017svb,Usman:2015kfa,Canton:2014ena,Cannon:2011vi,Cannon:2012zt} to extract
signals from data. To search for a wide range of parameters, a set of waveform templates is chosen carefully so that any potential signal would have high overlap with at least one of the
waveform templates, typically targeting no more than $10\%$ loss in detection rate~\cite{Sathyaprakash:1991mt,Babak:2006ty,Ajith:2012mn,Brown:2012qf,Capano:2016dsf}. This paper focuses on an improvement to the PyCBC analysis~\cite{Usman:2015kfa,Nitz:2017svb,pycbc-github} that was used to look for the gravitational waves from compact object mergers during the
second observing run of Advanced LIGO\footnote[1]{The modified ranking statistic described here was not introduced until after the analysis of GW170104}. The methods discussed in this paper can be extended for use with multiple gravitational-wave detectors, however, as Virgo data were not included in the initial matched-filtering based searches for compact binary mergers~\cite{Abbott:2017oio}, this paper focuses on the improvements gained in the analysis of LIGO data. The parameter space searched and the templates chosen for the O2 analysis are described in~\cite{DalCanton:2017ala}.

If the detector noise were Gaussian, matched filtering alone would be nearly sufficient to find signals within LIGO data, however, the data contains non-Gaussian 
noise transients which produce large signal-to-noise ratios~\cite{TheLIGOScientific:2016zmo,Nuttall:2015dqa}. In addition to the matched filter, the PyCBC analysis employs a signal consistency test~\cite{Allen:2005fk} as well as a consistency between the phase, amplitude, and time difference between the Hanford and Livingston observatories to rank gravitational wave candidates~\cite{Nitz:2017svb}. These candidates are compared to empirically measured background. The background is created by repeating the analysis after time shifting the data by an amount greater than the astrophysical time-of-flight~\cite{Usman:2015kfa}.

Part of the construction of the full ranking statistic, is the signal consistency re-weighted signal-to-noise, $\tilde{\rho}$~\cite{Babak:2012zx}, which measures the loudness of a signal in a single detector. In this paper we introduce a new signal consistency test, which targets a type of noise transient that is not well suppressed by existing methods, known as "blips"~\cite{TheLIGOScientific:2016zmo}. The mechanism for most instances of this
non-Gaussian noise source is not yet well understood. Many instances, however, contain excess power
at higher frequencies than expected for the waveform templates that these glitches are mistaken for. We also modify the single detector loudness measure and demonstrate that we are able to achieve $\sim 20\%$ improvement in the detection rate of binary black holes with masses greater than $60-100 M_{\odot}$.


\begin{figure}
\centering
\includegraphics[width=0.49\textwidth]{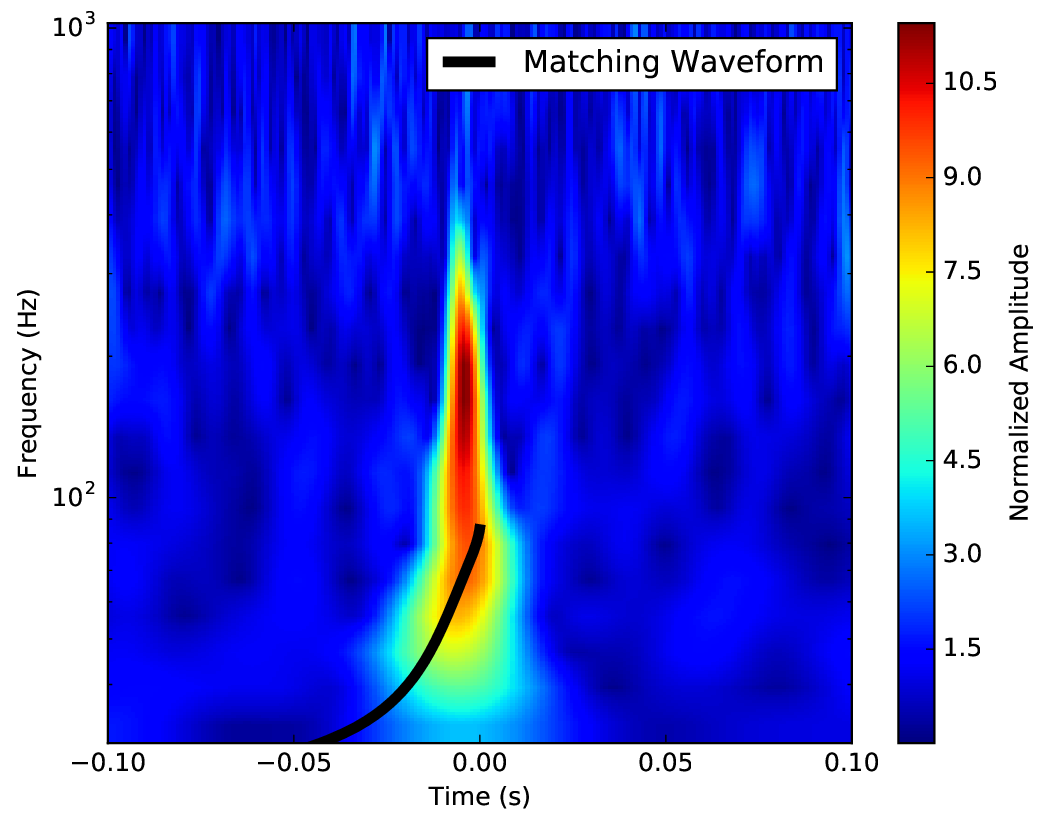}
\caption{The time-frequency representation of an example transient glitch in Hanford data. In black we have the time-frequency evolution of the best matching waveform. This is an example of the kind of waveform that current signal consistency tests are not well prepared to separate, as
the lower frequency content of the glitch matches the waveform template, even though the transient glitch clearly has excess power at higher frequencies contained in the waveform. The signal consistency test proposed in this paper is able to successfully downrank this event.
\label{f:glitch}}
\end{figure}

\section{Ranking Candidates and Signal Consistency}
\label{sec:regular}

Matched filter signal-to-noise ratio (SNR) is used in modeled searches for gravitational-waves from compact binary mergers~\cite{Usman:2015kfa,Nitz:2013mxa,Messick:2016aqy}. This has been shown to be optimal for extracting signal from Gaussian noise~\cite{Maggiore:1900zz,Creighton:2011zz}. When implicitly maximizing over an
unknown amplitude and phase of a potential signal, represented by a waveform template, $h$, this can be defined as

\begin{equation}
 \rho^2 \equiv \frac{\|\left\langle s | h \right\rangle\|^2}{\left\langle h | h \right\rangle}
\end{equation}
\label{eq:snr}

where the inner product is

\begin{equation}
 \langle a|b\rangle = 4 \int^\infty_0 \frac{\tilde{a}(f)\tilde{b}^*(f)}{S_n(f)} df
\end{equation}

and $s$, and $S_n(f)$ are the strain data and the estimated one-sided power spectral density of the noise around the time of a candidate event, respectively.

The detector noise, however, is not Gaussian nor stationary. A canonical method used since initial LIGO to discriminate between gravitational-wave signals from non-Gaussian transient noise, which make create large peaks in the SNR times series (a trigger), has been comparing the morphology of a candidate signal to the expectation from the triggering template waveform, $h$~\cite{Allen:2005fk,Usman:2015kfa}. We can construct this standard chi-squared~\cite{Allen:2005fk} test by sub-dividing this template waveform into p non-overlapping frequency bins. Each bin is constructed so that each contributes equally to the SNR of a perfectly matching signal. The number of bins has been empirically tuned by comparing the distribution of single detector background triggers produced from engineering data leading up to LIGO's second observing. Several formulas which determine the number of bins are tested and the value the gives a trigger distribution with the smallest excess at large statistic values is selected~\cite{TheLIGOScientific:2016pea,TheLIGOScientific:2016qqj}. We construct the chi-squared test as follows,
\begin{equation}
\chi^2_r = \frac{1}{2p-2}\sum_{i=1}^{i=p} \left\|\langle s|h_i \rangle - \langle h_i|h_i \rangle\right\|^2.
\end{equation}
If the data, $s$, is adequately described by Gaussian noise with an embedded signal that is well described by the waveform template $h$, this will follow a reduced $\chi^2$ distribution with $2p - 2$ degrees of freedom~\cite{Allen:2005fk}. Many classes of non-Gaussian noise have been demonstrated to take larger values~\cite{Allen:2005fk, Babak:2012zx}, creating separation between signals and noise artifacts. There are a number of different techniques for combining the $\chi^2$ test with the signal-to-noise ratio to produce a ranking statistic~\cite{Babak:2012zx,Adams:2015ulm,Messick:2016aqy,Nitz:2017svb}. In this paper we will focus on a modification to the re-weighted SNR, introduced in~\cite{Babak:2012zx}, which is a key component of the current ranking statistic used by the PyCBC analysis~\cite{Nitz:2017svb,Usman:2015kfa} method. This re-weighted SNR, $\tilde{\rho}$, is given as
\begin{equation}
 \tilde{\rho} = \begin{cases} 
        \rho & \mathrm{for}\ \chi^2_r <= 1 \\
        \rho\left[ \frac{1}{2} \left(1 + \left(\chi^2_r\right)^3\right)\right]^{-1/6} & 
        \mathrm{for}\ \chi^2_r > 1
    \end{cases},
\end{equation}
While $\tilde{\rho}$ has been shown to successfully remove most non-Gaussian transients from searches for low mass compact binary mergers ($M_{total} < 25 M_{\odot}$)~\cite{Babak:2012zx}, searches for higher mass mergers, where the observable portion of signals are typically fractions of a second in duration, have backgrounds more polluted by short duration transient noise~\cite{Abadie:2011kd,TheLIGOScientific:2016qqj}. 
Fig.~\ref{f:glitch} shows an example short duration noise transient that is able to fool the $\chi_r^2$ 
discriminator with $\chi^2_r= 1.1$, and so produces a trigger in the Hanford detector with high $\tilde{\rho}$. In the next section we construct a new signal consistency test to help distinguish this type of noise transient from a short duration gravitational-wave signal.

\begin{figure*}
  \centering
      \includegraphics[width=\textwidth]{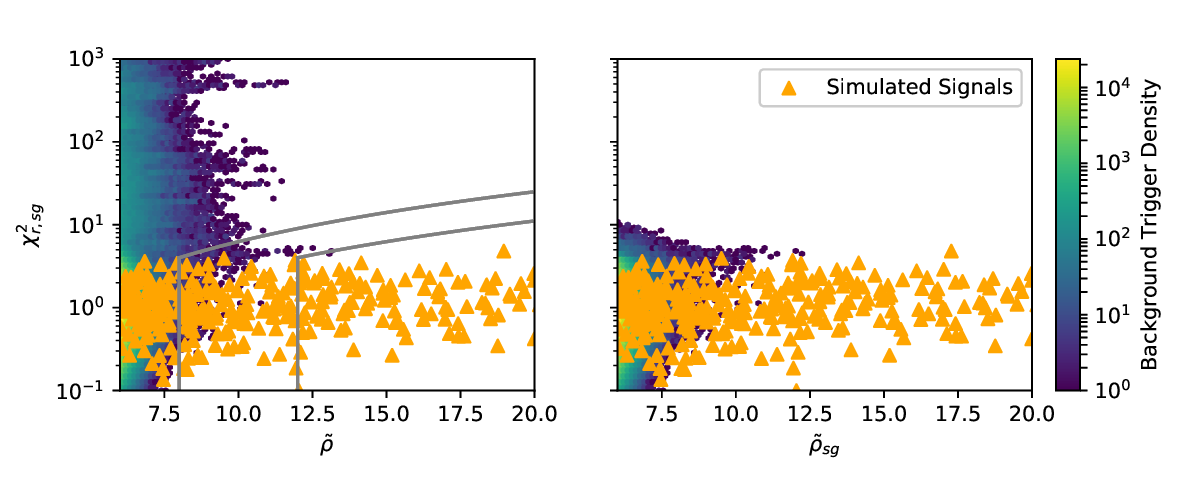}
  \caption{
  The distribution of background triggers and simulated signals (orange) as recovered from a representative portion of Hanford data from LIGO's second observing run. (Left) we show the distribution as a function of the previously used re-weighted SNR, $\tilde{\rho}$, and the high frequency, sine-Gaussian $\chi^2_{sg}$. The colorbar shows the number of single detector background trigger within each hexagonal bin. (Right) shows the same distribution as a function of $\tilde{\rho}_{sg}$. The two gray lines show a constant contour of $\tilde{\rho}_{sg}$, at 8 and 12, respectively. Only background from waveform templates with total mass greater than 40 solar 
masses is shown. We see that a large number of these single detector background triggers are clearly separated from the injection population, and are assigned a high $\tilde{\rho}$. The $\tilde{\rho}_{sg}$ statistic successfully downranks background triggers, while simultaneously leaving the vast majority of simulated signals at the same ranking statistic value. 
}\label{f:injrank}
\end{figure*}

\section{High Frequency Sine-Gaussian $\chi^2$ Discriminator}
\label{sec:sg}

As Fig.~\ref{f:glitch} demonstrates, some glitches which are not distinguishable from short duration
gravitational-wave signals using the $\tilde{\rho}$ statistic, have excess power at higher frequencies
than the best matching gravitational-wave model would actually contain. The aim is to distinguish this class of glitch by measuring this excess power. This is done by placing a series of sine-Gaussian tiles at the time of the candidate event, typically defined by the peak amplitude of the template waveform.  The tiles are placed at frequencies where we do not expect a true gravitational signal to contain power. 
A new signal discriminator, $\chi^2_{r,sg}$ can be written down as the sum of the signal-to-noise ratios squared of each individual sine-Gaussian tile.

\begin{equation}
\chi^2_{r,sg} \equiv \frac{1}{2N}\sum^N_i \rho_i^2 =  \frac{1}{2N}\sum^N_i <s|\tilde{g}_i(f, f_0, t_0, Q)>^2 
\end{equation}

where the expectation value of this reduced form is one, and will follow a reduced $\chi^2$ distribution with 2N degrees of freedom, when the data contains stationary, Gaussian noise at the location of each
tile. Each sine-Gaussian tile, $g_i$, can be defined in the time domain as

\begin{equation}
g \equiv A \exp\left(-4\pi f_0^2\frac{(t-t_0)^2}{Q^2} \right)\cos(2\pi f_0 t + \phi_0)
\end{equation}

where, $f_0$ and $t_0$, are the central frequency and time of the sine-Gaussian respectively, Q is the quality factor, A is the amplitude, and $\phi_0$ is the phase. The phase and amplitude are implicitly maximized over. As an overall amplitude factor of a template does not affect the SNR as defined in ~\ref{eq:snr}, we choose to set the amplitude A to one. It has been shown that some glitches in LIGO data can be approximately modeled as sine-Gaussian and the effect on matched filtering has been studied~\cite{Canton:2013joa,Blackburn:2008ah}.

The starting frequency of the first tile is determined by examining the expected contribution of power a signal would produce in different frequency bands. The current configuration places tiles from 30-120 Hz above the final frequency of a given template waveform, spaced in intervals of 15 Hz. A constant Q value of 20 has been selected. The frequency spacing and Q of the tiles was tested on a short subset of data, and the current values were empirically chosen from a limited number of hand selected variations. It may be possible to achieve a more optimal placement and choice of Q for each tile. This new discriminator, $\chi^2_{r,sg}$ is then combined with the re-weighted SNR described earlier to generate a new single detector test statistic, $\tilde{\rho}_{sg}$, which can be expressed as,

\begin{equation}
\label{eq:sg}
 \tilde{\rho}_{sg} = \begin{cases} 
        \tilde{\rho} & \mathrm{for}\ \chi^2_{r,sg} <= 4\\
        \tilde{\rho} (\chi^2_{r,sg} / 4)^{-1/2} & 
        \mathrm{for}\ \chi^2_{r,sg} > 4 
    \end{cases}.
\end{equation}

The re-weighted SNR, $\tilde{\rho}$, has the property that for candidates which are well approximated by one of our templates in we recover the standard matched filter signal-to-noise ratio. This same property is preserved for our new single detector loudness $\tilde{\rho}_{sg}$. This ansatz was chosen to allow for the expected variation in $\chi^2_{sg}$ in Gaussian noise. For values of $\chi^2_{sg}$ less than 4, we recover exactly the standard re-weighted SNR, $\tilde{\rho}$. Using the formula in Eq.~\ref{eq:sg} allows for some variation of the signal from our template waveforms, which may result in signal power spilling into the time-frequency region tested by the tiles.

\begin{figure}
\centering
\includegraphics[width=0.49\textwidth]{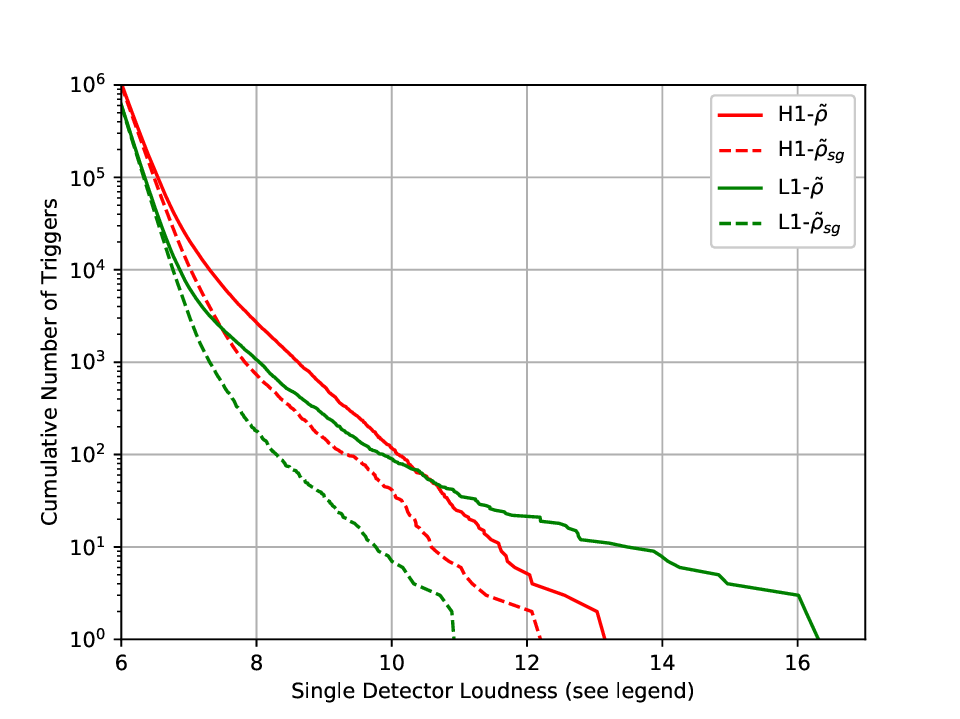}
\caption{The cumulative number of single detector triggers
as a function of ranking statistic in both the Hanford (red) and
Livingston (green) data during a representative time period for waveform templates
with total mass greater than 40 solar masses. The original, re-weighted SNR, $\tilde{\rho}$, is shown with solid lines
while the improved single detector loudness, $\tilde{\rho}_{sg}$, is shown with dotted lines. The new statistic
significantly reduces the number of triggers.}
\label{f:hist}
\end{figure}

The effect of this new single detector loudness metric, $\tilde{\rho}_{sg}$, is demonstrated in Fig.~\ref{f:injrank}, where we plot the distribution of background triggers from the Hanford data for the period from Feb 3rd to Feb 12th, 2017. We see that $\chi^2_{r,sg}$ provides additional information not encapsulated
in $\tilde{\rho}$. We have also overlaid a population of simulated signals, whose population is described in the next section. For most signals, the loudness statistic, $\tilde{\rho}$ or $\tilde{\rho}_{sg}$ will take the same value. Fig.~\ref{f:hist} shows the cumulative number of background triggers in the Hanford
and Livingston detectors above each statistic. In Figs.~\ref{f:injrank}-\ref{f:hist} we have restricted to showing the background from waveform templates
with total mass greater than 40 solar masses. At values between 7-10, we find an order of magnitude decrease
in the number of background triggers using $\tilde{\rho}_{sg}$, and factor of 3 decrease in Hanford. This decrease is the result of
the downranking visible in Fig.~\ref{f:injrank} of triggers with poor $\chi^2_{r, sg}$ values and directly leads to improved sensitivity by producing a cleaner background. This
will vary between analysis periods, but a large reduction is consistently observed. In the next section
we will estimate the impact this has on the sensitivity of the PyCBC analysis.

\section{Impact on Sensitivity}
\label{sec:test}

In this section we evaluate the impact on search sensitivity when using using the $\tilde{\rho}_{sg}$ statistic. So far, we have described the construction of an improved
single detector loudness, $\tilde{\rho}_{sg}$. To build a complete two-detector ranking statistic we use the methods described in~\cite{Nitz:2017svb}, where we have substituted $\tilde{\rho}$ with $\tilde{\rho}_{sg}$, to rank candidate coincident events and coincident background events.

We simulate an isotropic over the sky and volumetric over distance population of binary black hole mergers with a uniform distribution of component masses. A distance cutoff is imposed so that injections are not placed where their expected SNR would be significantly less than the minimum cutoff of the search (typically SNR $\sim 5.5$). The minimum component mass, $M_{1,2}$ is 2 solar masses, and the total mass, $M_{total}$ ranges from 10 to 100. Signals are approximated using the waveform model introduced in~\cite{Bohe:2016gbl}. The spin distribution was assumed to be aligned with the orbital angular momentum of the binary and the spin of each component black hole is uniformly distributed between $-0.998$ to $+0.998$. We use the same set of templates introduced for the production O2 analysis~\cite{DalCanton:2017ala}.

\begin{figure}
\centering
\includegraphics[width=0.49\textwidth]{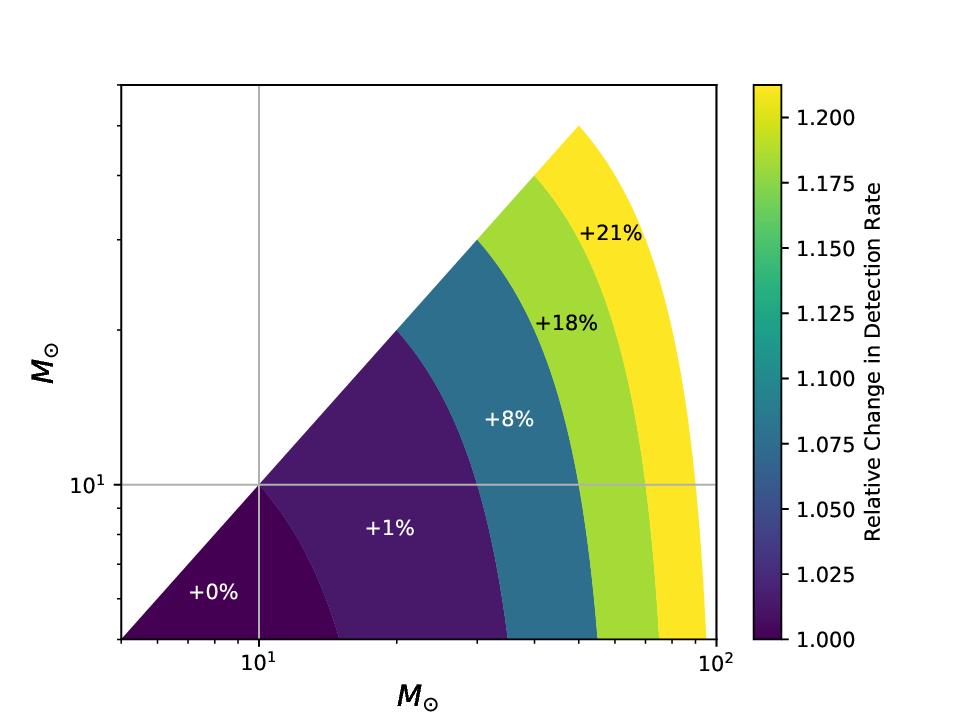}
\caption{The improvement in detection rate of binary black hole mergers
in a representative sample of data from the second observing run of Advanced LIGO. We see that at high masses, the improvement in detection rate is the highest.}
\label{f:vt}
\end{figure}

The PyCBC analysis~\cite{pycbc-github,Usman:2015kfa,Nitz:2017svb} is used to measure the background and significance of $\mathcal{O}(10^4)$ simulated signals. Each simulated merger has been added within the analysis to the LIGO data spanning Jan 4th - Feb 12th, 2017. Fig.~\ref{f:vt} shows the improvement in the rate of simulated detections using the modified ranking statistic. We find significant improvements in detection rate which increase with the total mass of the binary black hole. This is due to the fact that as the mass of the merger increases, signals are more difficult to distinguish from transient glitches, due to the increasing overlap in their time and frequency evolution. In addition, the sensitivity to lower mass mergers is unaffected. This is expected, as the detector noise does not regularly produce long duration transient glitches that could have the same time and frequency evolution as the signal from an astrophysical gravitational-wave merger.

\section{Conclusions}

In this work we have presented a novel method for ranking the single detector loudness of a 
gravitational-wave candidate trigger. This method takes into account the morphology of
a potential signal to look for excess power at higher frequencies than would be expected to be produced by a given candidate. We note that signals that are not well modeled by the templates, will be less effectively separated from background, so efforts to expand the space of models to include additional physical effects such as higher modes~\cite{Harry:2017weg}, and precession~\cite{Harry:2016ijz}, may aid in detection efficiency.

The method allows us to separate a large fraction of the short duration transient noise that is present in Hanford and Livingston data from the gravitational-waves expected from binary black hole mergers. This improves the overall sensitivity of the search by $\sim 20\%$ to gravitational-waves from black hole mergers with total mass $60-100 M_{\odot}$.

\label{sec:conclude}


\ack
We thank the LIGO Scientific Collaboration for access to the data and gratefully acknowledge the support of the United States National Science Foundation (NSF) for the construction and operation of the LIGO Laboratory and Advanced LIGO as well as the Science and Technology 
Facilities Council (STFC) of the United Kingdom, and the Max-Planck-Society 
(MPS) for support of the construction of Advanced LIGO. Additional support 
for Advanced LIGO was provided by the Australian Research Council. In addition, we acknowledge the Max Planck Gesellschaft for support and the Atlas cluster computing team at AEI Hannover where this analysis was carried out. We also thank Thomas Dent, Andrew Lundgren, Ian Harry, and Miriam
Cabero for useful discussions.

\section*{References}
\bibliographystyle{iopart-num.bst}
\bibliography{references.bib}

\end{document}